\documentstyle[11pt]{article}
\renewcommand{\theequation}{\arabic{equation}}
\newcommand{\be}{\begin{equation}}
\newcommand{\ee}{\end{equation}}
\newcommand{\bea}{\begin{array}}
\newcommand{\ea}{\end{array}}
\newcommand{\beqa}{\begin{eqnarray}}
\newcommand{\eeqa}{\end{eqnarray}}
\newcommand{\bean}{\begin{eqnarray*}}
\newcommand{\eean}{\end{eqnarray*}}

\def\up#1{\leavevmode \raise.16ex\hbox{#1}}

\newcommand{\gapproxeq}{\lower
 .7ex\hbox{$\;\stackrel{\textstyle >}{\sim}\;$}}
\newcommand{\lapproxeq}{\lower .7ex\hbox{$\;\stackrel
{\textstyle <}{\sim}\;$}}

\renewcommand{\theequation}{\thesection.\arabic{equation}}

\newcounter{appendice}
\newcommand{\appendice}
{
\setcounter{equation}{0}
\renewcommand{\theequation}{\Alph{appendice}.\arabic{equation}}
\addtocounter{appendice}{1}
{\bf Appendix \Alph{appendice}}
}

\def\thebibliography#1{{\bf REFERENCES\markboth
 {REFERENCES}{REFERENCES}}\list
 {[\arabic{enumi}]}{\settowidth\labelwidth{[#1]}\leftmargin\labelwidth
 \advance\leftmargin\labelsep
 \usecounter{enumi}}
 \def\newblock{\hskip .11em plus .33em minus -.07em}
 \sloppy
 \sfcode`\.=1000\relax}

\begin{document}


\centerline{ \LARGE  Generalized Coherent State Approach to Star Products}
\bigskip
\centerline{\LARGE and }
\bigskip
\centerline{ \LARGE  Applications to the Fuzzy Sphere }
\vskip 2cm

\centerline{ {\sc  G. Alexanian$^{a}$,  A. Pinzul$^{b}$ and A. Stern$^{b}$ }  }

\vskip 1cm
\begin{center}
{\it a)  Department of Physics, Syracuse University,\\ Syracuse, 
New York 13244-1130,  USA\\}
{\it b) Department of Physics, University of Alabama,\\
Tuscaloosa, Alabama 35487, USA}
\end{center}

\vskip 2cm

\vspace*{5mm}

\normalsize
\centerline{\bf ABSTRACT}
We construct a star product associated with an arbitrary two dimensional Poisson structure
 using  generalized
coherent states on the complex plane.  From our approach one easily recovers the
star product  for the fuzzy torus,
and also one for the fuzzy sphere.  For the latter we need to define the `fuzzy'
stereographic projection to the plane and the fuzzy sphere integration measure,
 which in the commutative limit reduce to
the usual formulae for the sphere.

\vskip 2cm
\vspace*{5mm}

\newpage
\scrollmode

\section{Introduction}

The star product is an important tool for deformation quantization and noncommutative
geometry.
 The most well studied star product is often referred to as the Moyal star product \cite{groe},\cite{moy}. 
(For a nice review see \cite{zac}.)
It allows for a quantum mechanical description on phase space.
 In recent times it has found application in the string theory approach to
noncommutative geometry.  It is of particular importance for the fuzzy torus
and has a simple form when acting on the periodic modes.   Another star product
  due to Grosse and Presnajder\cite{GP} is applicable to
the fuzzy sphere\cite{Mad},\cite{Wat},\cite{BGY},\cite{BBIV},\cite{BV},\cite{BMO}, and
 is also of current interest in string theory\cite{ARS}.  This star product
is constructed from coherent states on $S^2$ and is generalizable to arbitrary
coset manifolds.\cite{Perel}  By relying on coherent states, the property of
 associativity is assured.  The only other requirement on the star product
  is that it has a proper
commutative limit.  For this one assumes  it to be a function
of a parameter, say $\hbar$, which can be Taylor expanded about
 $\hbar =0$.  At zeroth order the star product reduces to the ordinary product, and
at first order the star (or Moyal) commutator should be proportional to the Poisson
bracket.  The relevant issue is to find the star product associated with a given
Poisson manifold.  In this regard, a nontrivial constructive approach was given by
 Kontsevich\cite{Kon} which is applicable for arbitrary Poisson structures.

The approach taken in this article 
 is along the lines of Berezin  quantization\cite{Brz},\cite{Perel}, and
relies on  generalized
coherent states on the complex plane developed by
 Man'ko, Marmo,  Sudarshan and Zaccaria\cite{mmsz}.  Associativity is once again assured,
and the results can  be applied to general two dimensional Poisson structures.    From  
our approach one easily recovers the Moyal star product (or more precisely, an equivalent construction due to
 Voros\cite{vor},\cite{zac}),
and also a star product for the fuzzy sphere.
The generalized coherent states have the usual property that they form an over complete
basis for an infinite dimensional Hilbert space.  They  are eigenstates of a `deformed'
 annihilation operator.   For the fuzzy sphere, we must perform
a `stereographic projection' of the
usual operators generating the fuzzy sphere algebra, and a slight modification of the above
 procedure is necessary.  This is because away
from the classical limit  a) the Hilbert space is finite dimensional,
 and a related issue is b) the  deformed annihilation
operator is not diagonalized  by the  coherent states.  The deformed annihilation
operator in this case is the operator analogue  of the stereographic coordinate.
Our result for the  star product of the  fuzzy sphere is expressed as an integral of
hypergeometric functions.
We show that it has
the proper commutative limit $j\rightarrow \infty$, where  $2j+1$ is 
 the number of dimensions ($j=$half-integer), and further that the fuzzy stereographic projection and
fuzzy integration measure  reduce to the
 usual formulae for the sphere in the limit.  As usual, the stereographic projection
has a coordinate singularity, which we choose at the north pole,
in the commutative limit.  However, away from $j\rightarrow \infty$,
 we can
argue that there is no coordinate singularity, simply because the
 quantum mechanical probability of being at the north pole is zero.
The fuzzy sphere star product of ref. \cite{GP}  is expressed in terms of the three {\it dependent}
coordinates of the sphere obtained from embedding it in $R^3$.
 Their product is later projected down to the plane
via the standard stereographic projection.   Ours, on the other hand, is written
directly on the two dimensional domain, and may  therefore be a more convenient tool
for writing down noncommuting field theories. Since the fuzzy sphere
is a (finite dimensional) matrix model,
 the corresponding field theory must be absent of any divergences, in contrast to
 the case
of field theories on the fuzzy torus.    We plan to pursue this in
future works.

In the next section we give a general formula for the star product in two dimensions based
on generalized coherent states.  There we show how to recover the Voros product.
In the  general case, we find a complete set of functions which close under the
action of the star product. In section 3, we apply the star product formula to the case
of the
fuzzy sphere and write down the fuzzy stereographic projection.   The appendix shows how
to express the normalization factor for the fuzzy sphere coherent states in terms of
a hypergeometric function.

\section{Star Product}

\subsection{General Properties}

For our star product we introduce generalized coherent states $| \zeta >  $ ,
 with the label $ \zeta $  corresponding
to a point on the complex plane.  They are assumed to
 form an (overcomplete) basis for Hilbert space $H$.
As is usual for coherent states, they are unit vectors $<\zeta|\zeta>=1$ and satisfy
the completeness relation
\be 1 = \int d\mu (\zeta,\bar\zeta )\;  | \zeta >  < \zeta |
\;, \label{comp}\ee where $ d\mu (\zeta,\bar\zeta )  $ is the appropriate measure
  on the complex plane, and the bar denotes complex conjugation.
We also assume the existence of another basis for $H$, and in terms of this basis
the states $| \zeta >  $ are expressible in  a
power series in $\zeta$ times some overall normalization.

To every operator $A$ on Hilbert space
$H$ one can associate a function ${\cal A}(\zeta,\bar \zeta )$
on the complex plane according to
\be {\cal A}(\zeta,\bar \zeta ) =<\zeta | A|\zeta > \ee
An associative product for two such functions is then defined by
\beqa {\cal A}(\zeta,\bar \zeta ) \star {\cal B}(\zeta,\bar \zeta ) &=&
<\zeta | AB|\zeta >  \cr
& & \cr
&=& \int d\mu (\eta,\bar\eta )\;
<\zeta | A
 | \eta >  < \eta |B|\zeta > \label{eq3} \eeqa  If $|\zeta>$  (  $<\zeta|$ ) is, up to
a normalization factor, analytic (anti-analytic) in $\zeta$, then the ratio
 $< \eta |A|\zeta >/< \eta |\zeta >    $
 is analytic in $\zeta$
and  anti-analytic in $\eta$.  Furthermore,
 it can be obtained from 
 $\;{\cal A}(\zeta,\bar \zeta )$ by acting with the translation operator twice
\beqa  e^{-\zeta \frac{\partial}{\partial \eta}} \;e^{\eta \frac{\partial}{\partial \zeta}}
\;{\cal A}(\zeta,\bar \zeta ) &=&  e^{-\zeta \frac{\partial}{\partial \eta}} \;
\;\frac{<\zeta | A  |\zeta + \eta > }{ <\zeta |\zeta +\eta >}  
=  \frac{<\zeta | A  | \eta > }{ <\zeta | \eta >}   \cr
 & &   \cr
e^{-\bar \zeta \frac{\partial}{\partial \bar \eta}} \;e^{\bar \eta \frac{\partial}{\partial \bar \zeta}}
\;{\cal A}(\zeta,\bar \zeta ) &=&  e^{-\bar \zeta \frac{\partial}{\partial  \bar \eta}} \;
\;\frac{<\zeta + \eta| A  |\zeta  > }{ <\zeta +\eta |\zeta >}  
=  \frac{<\eta | A  | \zeta > }{ <\eta | \zeta >}   
\eeqa
Alternatively, we can write $ e^{-\zeta \frac{\partial}{\partial \eta}} 
\;e^{\eta \frac{\partial}{\partial \zeta}}
$ (acting on $\eta$-independent functions)
 as an {\it ordered } exponential 
  $$  :\exp{  (\eta-\zeta ) \overrightarrow{ \frac\partial { \partial\zeta } } }: \;,$$
where the derivatives are ordered to the right in each term in the Taylor expansion, and 
they also act to the right.  Similarly, we define 
$$:\exp{ \overleftarrow{ \frac\partial{ \partial\zeta }}(\eta-\zeta)  }: \;,$$
where the derivatives are ordered to the left in each term in the Taylor expansion, and 
they  act to the left.
Substituting into (\ref{eq3}),
 we can then write the  product on functions of $\zeta$ and $\bar\zeta$ according to
\be    \star     = 
 \;\;\int d\mu (\eta,\bar\eta )\;   \;
:\exp{ \overleftarrow{ \frac\partial{ \partial\zeta }}(\eta-\zeta)  }:\;
\; |
<\zeta |\eta >  |^2\;     \;:
\exp{  (\bar\eta-\bar\zeta ) \overrightarrow{ \frac\partial { \partial\bar\zeta } } }:
\label{gspp}\ee
The product is thus determined once we know the measure
 $ d\mu (\zeta,\bar\zeta )  $ and the scalar product $ <\zeta |\eta >$.

The product (\ref{gspp})  has the property that  in general it is not symmetric.  It
 reduces to the ordinary product if the function on the right is analytic
in $\zeta$ and the function on the left is anti-analytic in $\zeta$
\be {\cal A}(\bar \zeta ) \star {\cal B}(\zeta ) ={\cal A}(\bar \zeta )
\; {\cal B}(\zeta ) \label{zbsz} \;.\ee
If we have that the states $|\zeta >$ are eigenvectors of some operator $\tilde {\bf a}$
with eigenvalues $\zeta$
\be \tilde {\bf  a}| \zeta > =\zeta |\zeta >\;,\label{eefa}\ee
then the  product (\ref{gspp}) between  two analytic functions also reduces to
the ordinary product
\beqa
 {\cal A}( \zeta ) \star {\cal B}(\zeta )
& =&
<\zeta|{\cal A}(\tilde{\bf a} )|\zeta> \star <\zeta|{\cal B}(\tilde{\bf a} )
|\zeta> \cr
&=&
<\zeta|{\cal A}(\tilde{\bf a} ){\cal B}(\tilde{\bf a} )|\zeta> \cr
&=& {\cal A}( \zeta )\; {\cal B}(\zeta ) \label{azbz} \eeqa
Similarly, then
 the  product  (\ref{gspp})  between  two anti-analytic functions  reduces to
the ordinary product
\be
 {\cal A}(\bar \zeta ) \star {\cal B}(\bar \zeta )
=  {\cal A}(\bar \zeta ) \; {\cal B}(\bar\zeta ) \label{absbb}
\ee
For  this we only need $ <\zeta |\tilde {\bf  a}^\dagger =\bar\zeta <\zeta | $ .  To get
nontrivial results we can take
 the function on the right to be anti-analytic
in $\zeta$ and the function on the left to be analytic in $\zeta$
\beqa
 {\cal A}( \zeta ) \star {\cal B}(\bar \zeta )
& =&
   <\zeta |{\cal A}( \tilde{\bf a} )
  \; {\cal B}(\tilde{\bf a}^\dagger )|\zeta> \cr
&=&  {\cal A}( \zeta ) \; {\cal B}(\bar \zeta )
\;+\;   <\zeta |\;[\;{\cal A}( \tilde{\bf a} )\;
,\; {\cal B}(\tilde{\bf a}^\dagger )\;]\;|\zeta> \label{axsbbz}
 \eeqa
which  can be evaluated once we know the commutation relations
for $\tilde {\bf a}$ and $\tilde {\bf a}^\dagger$.  Say the commutation relations are
of the form \be [\tilde{\bf a},\tilde{\bf a}^\dagger]=\;
F(\tilde {\bf a}\tilde
{\bf a}^\dagger)
\label{capaad}\ee for some function $F$, and $F$ can be Taylor expanded in some (commuting)
parameter $\hbar$, where the lowest order term is linear in $\hbar$.  For
the ``classical limit"  defined as $\hbar\rightarrow 0 $ , one demands that  \beqa
{\cal A}( \zeta ,\bar \zeta) \star {\cal B}(\zeta,\bar \zeta )
 &\rightarrow &{\cal A}( \zeta,\bar \zeta ) \;
 {\cal B}(\zeta, \bar \zeta ) \;,\\
{\cal A}( \zeta ,\bar\zeta) \star {\cal B}(\zeta, \bar \zeta ) -
{\cal B}(\zeta,\bar \zeta )\star {\cal A}( \zeta ,\bar\zeta)
&\rightarrow & O(\hbar )\;,\label{posbr} \eeqa
and the coefficient on the right hand side of (\ref{posbr}) is identified with
the Poisson bracket. [For the star product of an analytic function with an anti-analytic
function, these two conditions reduce to one:
${\cal A}( \zeta ) \star {\cal B}(\bar \zeta )\rightarrow {\cal A}( \zeta ) \;
 {\cal B}(\bar \zeta ) + O(\hbar ) $ thanks to (\ref{zbsz}) .]
  We thereby obtain all the properties of the star product.

\subsection{Standard Coherent States}

From the standard coherent states it is easy to recover the  Moyal\cite{groe}, \cite{moy}
(or actually the Voros\cite{vor}) star product. \footnote{After posting an earlier
version of this article on the hep-th archive, we were informed of a similar
 discussion  in   \cite{Read}.}
Here we identify $\tilde {\bf a}$ and $\tilde {\bf a}^\dagger $ with the
 standard lowering and
raising operators for the harmonic oscillator
 $ {\bf a}$ and $ {\bf a}^\dagger $, satisfying
\be [ {\bf a},
 {\bf a}^\dagger]=1\;.\label{aadag} \ee  (For the moment we supress $\hbar$.)
Then products such as
(\ref{axsbbz}) are easy to compute.
It is also easy to perform
 the integral in (\ref{gspp}) in this case.
The scalar product squared is
$ |<\zeta |\eta >   |^2 = e^{-|\zeta - \eta|^2} $
and the measure is \be d\mu (\eta,\bar\eta )\;   =\frac 1\pi d\eta_R \;d\eta_I \;,
\label{imscs}\ee
$\eta_R$ and $\eta_I $ being the real and imaginary parts of $\eta$.  The integrand
 in (\ref{gsp}) is then a Gaussian:
\beqa \star & =& \frac 1\pi \int d\eta_R \; d\eta_I \;  \;
:\exp{ \overleftarrow{ \frac\partial{ \partial\zeta }}\;(\eta-\zeta)  }:\;\;
\exp{(-|\zeta - \eta|^2)}\; \;:
 \exp{  (\bar\eta-\bar\zeta)\;\overrightarrow{ \frac\partial{ \partial\bar\zeta }}}:\cr
& =&\frac 1\pi  \int  d\eta_R \;d\eta_I \;  \;
\exp{ \overleftarrow{ {\partial\over{ \partial\zeta }}}\;\eta  }\;\;
\exp{(-| \eta|^2)}\;\;
 \exp{  \bar\eta \;\overrightarrow{ \frac\partial{ \partial\bar\zeta  }}}\cr
&=&\;\;
\exp{ \overleftarrow{ \frac\partial{ \partial\zeta }}\;\;
\overrightarrow{ {\partial\over{ \partial\bar\zeta} }}}\eeqa
Note that the ordering of the exponential function can be dropped after
the change of integration variables.
For this to be a star product it must contain  a parameter (say $\hbar$) which
admits a commutative limit ($\hbar\rightarrow 0$).  This is easily done by a
rescaling of the coordinates
$\zeta \rightarrow \frac1{\sqrt{ \hbar}}\zeta$ and then  $$\star\;\;\rightarrow\;\;
\exp{\;\hbar \;\overleftarrow{ \frac\partial{ \partial\zeta }}\;\;
\overrightarrow{ {\partial\over{ \partial\bar\zeta} }}}$$
This is the star product of Voros\cite{vor}, which is equivalent to the standard star
product on the plane\cite{groe}, namely  $$\tilde\star\;\;=\;\;
\exp{\;\frac{\hbar}2 \;\biggl(\overleftarrow{ \frac\partial{ \partial\zeta }}\;\;
\overrightarrow{ {\partial\over{ \partial\bar\zeta} }}\;-\;
\overleftarrow{ \frac\partial{ \partial\bar\zeta }}\;\;
\overrightarrow{ {\partial\over{ \partial\zeta} }}\biggr)}$$
The equivalence relation is $T ( {\cal A}) \star T ( {\cal B})
=T ( {\cal A}\; \tilde\star \; {\cal B}) \;,$ where $T$ is a nonsingular operator.
Using the identity  $$\exp{\;\frac{\hbar}2 \;{ \frac\partial{ \partial\zeta }}\;
{ {\partial\over{ \partial\bar\zeta} }}}\;\;({\cal A}\; {\cal B})=
{\cal A}\; \exp{\;\frac{\hbar}2 \;\biggl(\overleftarrow{ \frac\partial{ \partial\zeta }}\;\;
\overleftarrow{ {\partial\over{ \partial\bar\zeta} }}\;+\;
\overrightarrow{ \frac\partial{ \partial\bar\zeta }}\;\;
\overrightarrow{ {\partial\over{ \partial\zeta} }}+\;\overleftarrow{ \frac\partial{ \partial\zeta }}\;\;
\overrightarrow{ {\partial\over{ \partial\bar\zeta} }}\;+\;
\overleftarrow{ \frac\partial{ \partial\bar\zeta }}\;\;
\overrightarrow{ {\partial\over{ \partial\zeta} }}\biggr)}
\; {\cal B}
 $$ it can be easily checked that the relevant operator is\cite{zac}
$$T=\exp{\;\frac{\hbar}2 \;{ \frac\partial{ \partial\zeta }}\;\;
{ {\partial\over{ \partial\bar\zeta} }}}  \;.$$
(An alternative connection between coherent states and the Moyal
star product is deduced in \cite{man2},\cite{zac2}.)

\subsection{Deformed Coherent States}

A more  general class of coherent states on the complex plane was given in
\cite{mmsz}.   These coherent states provide a more convenient basis than
the standard coherent states
when studying functions of operators  
 $\tilde {\bf a}$ and $\tilde{\bf a}^\dagger $.
 Now we assume such operators satisfy the general
commutation relations (\ref{capaad}), while the coherent states satisfy
(\ref{eefa}).  The expectation values of $\tilde {\bf a}$ and
 $\tilde{\bf a}^\dagger $ for the coherent state  $|\zeta>$ are $\zeta$ and
$\bar\zeta$, respectively, and the star product  (\ref{gspp})  can be directly
applied to functions of these variables.

The procedure of  \cite{mmsz} 
requires the existence of a  map from the usual
 harmonic oscillator algebra generated by annihilation and creation operators
 ${\bf a}$ and ${\bf a}^\dagger$, satisfying (\ref{aadag}),
 to the  algebra generated by $\tilde {\bf a}$ and $\tilde{\bf a}^\dagger $.
The  map is
 expressed in
the form  \be \tilde {\bf a}= f({\bf  n}+1)\; {\bf a }\;,\label{atfa}\ee
  ${\bf  n}$ being
 the number operator ${\bf  n}={\bf a}^\dagger {\bf a}$,
 and the function $f$ being
determined from $F$.  In this section
we regard $f({\bf n})$ as a nonsingular function,
while we adapt the formalism to a singular function in the section 3.
Following \cite{mmsz} we restrict to real functions, as only the real 
part of $f$ determines $F$. 
 We can  introduce the Hilbert space ${\tt H}$
spanned by
orthonormal states $|n> $, $n=0,1,2,...$, with
 ${\bf a}|0>=0$ and ${\bf n}|n>=n|n>$.
Following \cite{mmsz}  one can  construct the analogue of the standard coherent states
 according to:
\beqa |\zeta > &=& 
 N (|\zeta |^2  )^{-\frac12}\; \exp{\{\zeta f({\bf n})^{-1}{\bf a}^\dagger\}}\;
f({\bf n})^{-1} \;|0> \cr  & &\cr  &=& 
  N (|\zeta |^2  )^{-\frac12}\;\sum_{n=0}^{\infty} \frac{\zeta^n }{\sqrt{n!}\;
[f(n)]!} \;|n> \;,\label{ndcs}\eeqa
where  $[f(n)]!= f(n)f(n-1)...f(0)$.   These states
 are  diagonal in $\tilde {\bf a}$,  rather than ${\bf a}$,
 with associated eigenvalues $\zeta$ as in
(\ref{eefa}).
 Requiring them  to be of unit norm fixes $N (|\zeta |^2  )  $,
\be N (x  )  =  \sum_{n=0}^{\infty} \frac{x^{n} }{n!\;([f(n)]!)^2} \;, \ee
which reduces to the exponential function for standard coherent states.  
As with the standard coherent states, the states (\ref{ndcs}) are  not orthonormal.  Now
 \be<\eta | \zeta >  =   N (|\eta |^2  )^{-\frac12} \; N (|\zeta |^2  )^{-\frac12}
 N(\bar\eta\zeta)   \label{zetet} \ee
Substituting into  (\ref{gspp})  then gives
\be    \star     =
 \;\;\int d\mu (\eta,\bar\eta )\;   \;
:\exp{ \overleftarrow{ \frac\partial{ \partial\zeta }}(\eta-\zeta)  }:
\; \frac { N(\bar\eta\zeta) \;  N(\bar\zeta\eta)  }
  { N (|\eta |^2  ) \; N (|\zeta |^2  )}   \;     \;
:\exp{  (\bar\eta-\bar\zeta ) \overrightarrow{ \frac\partial { \partial\bar\zeta } } }:
\label{gsp}\ee

 From (\ref{zetet}) and the completeness relation, the
 integration measure
should satisfy  \be N(\bar \zeta \lambda )=
\int d\mu (\eta,\bar\eta )\;   \;
 \frac { N(\bar\zeta\eta) \;  N(\bar\eta\lambda)  }
  { N (|\eta |^2  )   }  \;, \label{zetlam} \ee  for arbitrary complex coordinates $\zeta$
and $\lambda$.
 If we assume that it is of the form  $ d\mu (\zeta,\bar\zeta )  = i
h (|\zeta |^2  ) \;d\zeta \wedge d\bar\zeta$ ,
 then the conditions on $h (|\zeta |^2  ) $ are
\be  \int_0^\infty \frac{d\rho \;\rho^{2n+1}  \; h (\rho ^2  )}{ N (\rho^2  )  }
 = \frac{n!}{4\pi}
([f(n)]!)^2  \;, \quad {\rm for}\;{\rm all }\;{\rm integer}\; n\ge 0.\label{coh}\ee 
Upon defining $g(x)=h(x)/N(x)$ , we can rewrite (\ref{coh}) as
\be  \int_0^\infty dx \; x^{s-1} g(x)  = \frac{\Gamma (s)}{2\pi}
([f(s-1)]!)^2  \;, \quad {\rm for}\;{\rm all}\; {\rm integer}\; s\ge 1\;.\label{sfh}\ee
By definition the right hand side is the Mellin transformation of $g(x)$.
 Then $g(x)$ can be written as inverse Mellin integral transform 
\be
g(x)=\frac{1}{2\pi i}\int^{c+i\infty}_{c-i\infty}\;\frac{\Gamma (s)}{2\pi}
([f(s-1)]!)^2\;x^{-s}  \;ds\;\;,\label{mellin}
\ee where here $s$ is treated as a complex integration variable (with some
 possible restrictions) and it is assumed that the function $[f(s-1)]!$ can be 
extended appropriately
over the entire integration region. 
For the case of the standard coherent states where f(s)=1, we then get
 that $g(x)={e^{-x}}/{2\pi}$. Taking into account that for
 standard case $N(x)=e^x$ we recover the integration measure (\ref{imscs}).

It is possible to compute the star product for a class of function
 without using a specific expression for measure. For example,
  $ 1 \star 1 = 1  \;,$ which follows from (\ref{zetlam}).
Furthermore,  using (\ref{coh}), for the functions
\be
{\cal Z}  _{nm} =\frac{{\bar\zeta }^n \zeta^m }{ N (|\zeta |^2  )} ,
\quad n,m=0,1,2,3,...\;,\label{calZ} \ee with the property $\bar{\cal Z}  _{nm} =
{\cal Z}  _{mn}$ ,
the unit function
acts as the identity with respect to the the star product:
$   1 \star {\cal Z}  _{nm}  =  {\cal Z}  _{nm}  \star 1 ={\cal Z}  _{nm}
  \;,$ and these functions form a closed algebra:
\be {\cal Z}  _{nm}  \star {\cal Z}  _{rs} ={m!\;([f(m)]!)^2}  \;\delta_{m,r} \;{\cal Z}  _{ns}
\label{fnfn}\ee   This algebra has the projectors
\be P_m = \frac{ {\cal Z}  _{mm}  }{{m!\;([f(m)]!)^2}  } \label{proj} \ee
The square (using $\star$) of all other functions vanishes.
We can write $\zeta$ and $\bar \zeta$ in terms of of functions $ {\cal Z}  _{nm}  $
\be \zeta = \sum_{n=0}^{\infty} \frac{ {\cal Z}  _{n\;n+1}   }{n!\;([f(n)]!)^2}\;,
\qquad \bar\zeta = \sum_{n=0}^{\infty} \frac{ {\cal Z}  _{n+1\;n}   }{n!\;([f(n)]!)^2}
\;,\label{bzaz}\ee
and thereby compute their star products:
\beqa \bar\zeta \star \zeta &=&|\zeta|^2 \label{bzsz1}\\ \zeta \star \zeta &=&\zeta ^2
\label{zsz1}
\\ \bar\zeta \star \bar \zeta &=&\bar \zeta ^2 \label{bzsbz1}\\ \zeta \star \bar\zeta &=&
|\zeta|^2 + <\zeta|F(\tilde {\bf a}\tilde
{\bf a}^\dagger) |\zeta > \;,\label{zsbs} \eeqa
 $$  <\zeta|F(\tilde {\bf a}\tilde
{\bf a}^\dagger) |\zeta >= \sum_{n=0}^{\infty}\biggl((n+1)f(n+1)^2-nf(n)^2\biggr) \; P_n
 \;\;,
$$ which is consistent with  (\ref{zbsz}-\ref{axsbbz}).

\subsection{q-oscillators}

Coherent states for q-oscillators were studied by many authors
 \cite{nel},\cite{qcs},\cite{klsc}.
The q-oscillators algebra is generated by $\tilde{\bf a}$, $\tilde{\bf a}^\dagger$ and
${\bf n}$ satisfying the algebra\cite{bied},\cite{mcfa}
$$
 \tilde{\bf a}\tilde{\bf a}^\dagger -q^{-1}\tilde{\bf a}^\dagger\tilde{\bf a}
= q^{\bf n}$$    \be [{\bf n}  , \tilde{\bf a} ] = -\tilde{\bf a}\;,\qquad
[{\bf n}  , \tilde{\bf a}^\dagger ] = \tilde{\bf a}^\dagger
\;,\ee where  we consider $q$ to be a real parameter,
and when $q\rightarrow 1$ we recover the ordianry oscillator algebra.
As the notation implies ${\bf n} $ can be identified with the `undeformed' number operator
${\bf  n}={\bf a}^\dagger {\bf a}$, while $f$ in (\ref{atfa}) is given by
\be f(n)^2 = \frac 1n  \;\frac{q^{n}-q^{-n}}
{q -q^{-1}}  \equiv \frac{[n]}{n}\;,\ee and   $N (x  )  $ corresponds to a q-exponential function,
\be  N (x  )  =  \sum_{n=0}^{\infty} \frac{x^{n} }{[n]!}\equiv e_q(x)
 \;,\ee $[n]!= [n][n-1]...[0]$.  According to \cite{nel}, for the measure one can use
\be h(\rho^2) d\rho^2 = \frac1{2\pi} e_q(\rho^2) e_q(-\rho^2) d_q\rho^2  \;,\ee 
where the `q-integration' over $\rho^2$ is defined by
\be \int_0^\varsigma g(x) d_qx = \varsigma (q^{-1/2} -q^{1/2}) \sum_{n=0}^{\infty} q^{n+1/2} 
g(\varsigma q^{n+1/2} ) \;,\ee
and $-\varsigma $ is the largest zero of $e_q(x)$.

\section{Fuzzy sphere}

With a small modification of the above procedure we can write
down the star product for  the fuzzy sphere.
The modification is necessary because we will no longer have (\ref{eefa}),
except in the commutative limit.  We associate deformed   annihilation
and creation operators,  $\tilde {\bf a}$ and $\tilde{\bf a}^\dagger $,
with the operator analogue of the  stereographic coordinates of a sphere.
Its algebra now leads to a highest weight state $|2j>$ and therefore finite ($2j+1$) 
dimensional
representations, which is another departure from the treatment of the previous
section.  It requires that we terminate the series in (\ref{ndcs}) and what follows,
and quantities computed previously now depend  on $j$.
The infinite series is recovered when $j\rightarrow \infty$, which is the commutative
limit of the fuzzy sphere.  For finite $j$, we are able to obtain  exact expressions
for the normalization factor   $N_j (|\zeta |^2  )$ and integration measure
$d\mu_j (\zeta,\bar\zeta ) $ in terms of hypergeometric functions.

\subsection{Fuzzy Stereographic Projection}

We first recall that
  the  stereographic  projection of a sphere of radius $1$,
$ x_ix_i=1 ,\;i=1,2,3,\;$ to  the complex plane which maps
the north pole to infinity
is given by  \be z =\frac { x_1-i x_2}{1-x_3} \;,\qquad
\bar z=\frac { x_1+i x_2}{1-x_3} \label{csp}\ee
To obtain the algebra of the fuzzy sphere one promotes the coordinates $x_i$ to operators
${\bf x}_i$'s ,
 satisfying commutation relations:
\be [{\bf x}_i,{\bf x}_j] = {i \alpha}\;\epsilon_{ijk}{\bf x}_k\;,\label{xixj}\ee
as well as   $ {\bf x}_i{\bf x}_i=1$, 
where $\alpha$ is a parameter which vanishes
in the commutative limit and $1$ now denotes the unit operator.  For $ \alpha ={1\over
{ \sqrt{j(j+1)}}}\;, \; j={1\over 2}, 1, {3\over 2},...$ , ${\bf x}_i$
 has finite dimensional representations,
which are simply given by ${\bf x}_i=\alpha {\bf J}_i$,
 ${\bf J}_i$ being the angular momentum matrices.
To define an operator analogue of the stereographic projection of operators ${\bf x}_i$ ,
 we need to choose an
ordering in the definition of  operators ${\bf z}$ and ${\bf z}^\dagger$.
  We define the following deformation map of the algebra:
\be {\bf z} =({\bf x}_1-i {\bf x}_2)(1-{\bf x}_3 )^{-1} \;,\qquad
   {\bf z}^\dagger =(1-{\bf x}_3 )^{-1} ({\bf x}_1+i {\bf x}_2) \label{fzysp} \ee
From  the commutation relations (\ref{xixj}) one gets
$$[{\bf z},\chi^{-1} ] =-{\alpha\over {2}} {\bf z }\;, \qquad [ {\bf z}^\dagger,\chi^{-1} ] ={\alpha\over {2}}  {\bf z}^\dagger $$
where  $\chi^{-1} ={1\over 2} (1-{\bf x}_3)$. It then follows that $\chi^{-1} $
commutes with $|{\bf z}|^2={\bf z}  {\bf z}^\dagger $ and
\be [{\bf z},{\bf  z} ^\dagger ] = \alpha \chi \biggl(1+|{\bf z}|^2 -{1\over 2} \chi (1+{\alpha\over 2}|
{\bf z}|^2 ) \biggr)  \;.\label{zzdag}  \ee   This is the analogue of eq. (\ref{capaad}),
the right hand side corresponding to the function $F$.
    For all finite dimensional matrix representations of the fuzzy sphere,
 $\chi^{-1}$ is represented by a nonsingular matrix, and the above equation makes sense.
 More generally, $\chi^{-1}$ is a nonsingular operator.   Since it is a
 hermitian  operator, and it should be possible to write it
in terms of $|{\bf z}|^2$.  To get its form  start with the identity
$$ {\bf z}\chi^{-2} {\bf z }^\dagger + \chi^{-1}  {\bf z }^\dagger {\bf z}
 \chi^{-1} +2 \chi^{-1}(\chi^{-1} - 1) = 0  $$
which follows from ${\bf x}_i{\bf x}_i=1$,
and apply the above commutation relations to get
$${\alpha \over 4}\xi\chi^2 - \chi (\xi +{\alpha\over 2})
+ 1+|{\bf z}|^2 = 0 \;,\qquad \xi=1+\alpha |{\bf z}|^2$$
having the solution
$${\alpha \over 2}  \chi = 1 +{\alpha \over {2\xi}}  -\sqrt{ {1\over \xi}+\biggl(
{\alpha \over{2\xi}}\biggr)^2}   $$
The sign choice is so that $\chi $ reduces to ${1+|{\bf z}|^2}$ in the  limit
 $j\rightarrow \infty$,
and thus the right hand side of (\ref{zzdag}) goes to
\be \frac1{2j} (1+|{\bf z}|^2)^2  \;.\label{lorhs}\ee
Since  for finite $j$, the eigenvalues of $(1 +{\alpha \over {2\xi}}  )^2$ are  greater than those of
$ {1\over \xi}+\biggl(
{\alpha \over{2\xi}}\biggr)^2$,
it follows that $\chi$ is an
invertable operator.  There is thus a $1-1$ correspondence between
representations of the algebra generated by ${\bf z}$ and
 $ {\bf z}^\dagger $ and the algebra of the fuzzy sphere.

One of the attractive features of the fuzzy sphere as a noncommutative space
 is that it is covariant with respect
to the same symmetry as the standard sphere, namely $SO(3)$.  (This is in contrast to
the case of the quantum sphere.\cite{Pod})  Upon stereographically
 projecting the symmetry transformations of the standard sphere to the complex plane
 one gets elements of the Mobius group.   These are nonlinear transformations, so in
 the case of the fuzzy sphere, we have to
be concerned with  operator orderings.  For  infinitesimal rotations of the fuzzy sphere,
$\delta{\bf x}_i  = \epsilon_{ijk} \epsilon_j{\bf x}_k$,  $\epsilon_j$ being an
infinitesimal c-number, ${\bf z}$ and ${\bf z}^\dagger$  undergo the variations
\beqa  \delta{\bf z} &=& -i\epsilon_3 {\bf z} -\frac{\epsilon_-}2 {\bf z}^2 +\frac{\epsilon_+}2
(\chi - 2 -\frac14 {\bf z}\chi {\bf z}^\dagger \chi ) \;, \cr & & \cr
 \delta{\bf z}^\dagger &=& i\epsilon_3 {\bf z}^\dagger -\frac{\epsilon_+}2 
{\bf z^\dagger}^2 +\frac{\epsilon_-}2
(\chi - 2 -\frac14 \chi {\bf z}\chi {\bf z}^\dagger  ) \;, \eeqa
 where
 $\epsilon_{\pm}=\epsilon _2\pm i\epsilon_1 $ . 

For the two, three and four dimensional representations of ${\bf z}$ and $|{\bf z}|^2$,
 we get, respectively,

$$ {\bf z}=\pmatrix{ 0 & 0 \cr 1+\sqrt{3} & 0 \cr} $$
\be |{\bf z}|^2 = {\rm diag}\biggl(\; 0\; ,\; 2(2+\sqrt{3})\;\biggr) \label{twod} \ee

$$ {\bf z}=\pmatrix{ 0 & 0 & 0 \cr 2+\sqrt{2} & 0 & 0 \cr 0 & 1 & 0 \cr}  $$
\be |{\bf z}|^2 ={\rm diag} \biggl(\; 0 \;,\; 2(3 + 2 \sqrt{2})\;,\;1\;\biggr)
 \label{tyhrd} \ee

$$  {\bf z}=\pmatrix{0 & 0 & 0 & 0 \cr \sqrt{3} + \sqrt{5} & 0 & 0 & 0 \cr
0 &    { 2\over 7} (1 + \sqrt{15}) & 0  & 0 \cr
 0 & 0 & {1\over 7} (  3 \sqrt{5} - \sqrt{3}) & 0 \cr }$$
\be|{\bf z}|^2 = {\rm diag}\biggl(\;0 \;,\; 2 (4 + \sqrt{15})\;,\; {8 \over 49}
         (8 + \sqrt{15})  \;,\; {6 \over 49}  (8 - \sqrt{15} )\;\biggr) \label{fourd}\ee
More generally, we
 denote the states of an irreducible representation $\Gamma^j$ as usual by $|j,m>$,
  $j=\frac 12, 1, \frac 32,...$,
 $m=-j,-j+1,...,j$.  The states span  vector space $H^j$.
Then $|{\bf z}|^2|j,m> =\lambda_{j,m}|j,m>$, with
\be \lambda_{j,m}=\frac{j(j+1) - m(m+1)}{(\sqrt{j(j+1)} - m -1)^2}\;.\label{lajm}\ee
  As $j\rightarrow \infty$, $\lambda_{j,m}$ ranges between $0$ and $8j+4$.

\subsection{Coherent States and Star Product}

We next define the  map from the  harmonic oscillator algebra.
This is clearly a singular map since $H^j$ is finite dimensional and the Hilbert space
 ${\tt H}$ for the latter is not.
  For irreducible representation $\Gamma^j$, we can restrict the map to act on
the finite dimensional
subspace of ${\tt H}$ spanned by the first $2j+1$ states $|n>  $, $n=0,1,2,...,2j$ .
More precisely,  we identify
$|j,m>$ in $H^j$ with $|j+m>  $ of ${\tt H}$ , and the  map is applied to this subspace.
Because the map depends on $j$ we include a $j$ subscript
 \be {\bf z}= f_j({\bf  n}+1)\; {\bf a } \label{mzta}\ee
 From (\ref{zzdag}) and (\ref{aadag})  the eigenvalues of $f_j({\bf  n}) ^2$ in
$H^j$ are $\lambda_{j,n-j-1}/n$.  Therefore,
 \be f_j({\bf  n}) = \frac{\sqrt{2j-{\bf  n }+1}}{\sqrt{j(j+1)} + j  - {\bf  n}}
  \label{fjn}  \ee
It is zero when acting on $|2j+1>$ and hence ${\bf z}^\dagger|2j> =0$ .  It is
ill-defined for harmonic oscillator states with $n>2j+1$.  The map (\ref{mzta}) is similar
to that of Holstein and Primakoff\cite{hp}, who instead go from the
angular momentum algebra to the oscillator algebra.

We now construct  coherent states, as before, with a linear combination of
${\bf  n}$ eigenstates.  Only here  we need to truncate the series at $n=2j$:
\be |\zeta,j> =  N_j (|\zeta |^2  )^{-\frac12}\;\sum_{n=0}^{2j} \frac{\zeta^n }{\sqrt{n!}\;
[f_j(n)]!} \;|n> \;.\label{trcs}\ee
The normalization condition is now
\be N_j (x  )  =  \sum_{n=0}^{2j} \frac{x^n} {n!\;([f_j(n)]!)^2} \;,\label{Nsj}\ee
which can be expressed in terms of a hypergeometric function [see appendix]\be   N_j (x  )
 =\;\frac {\Gamma(\gamma+2j+ 1)\;{}^2}
{(2j+1)!\;(2j)!\;\Gamma(\gamma)\;{}^2} \;\;
 {}_3F_2 (1,1,-2j;\gamma,\gamma; -x^{-1}) \;x^{2j} \;, \label{ceffj}
\ee
where $\gamma = \sqrt{j(j+1)} - j $.  To calculate the integration measure according to 
the general formula
(\ref{mellin}) we use the Mellin-Barnes type integral representation of
hypergeometric function ${}_2F_1$
$$
\frac{\Gamma (a)\Gamma (b)}{\Gamma (c)}\;{}_2F_1(a,b;c;-z)=\frac{1}{2\pi
  i}\int^{i\infty}_{-i\infty}\frac{\Gamma (a-s)\Gamma (b-s)\Gamma
  (s)}{\Gamma (c-s)}\;z^{-s}  \;ds\;\;, \qquad |\mbox{arg}(z)|<\pi\;,
$$
where the path of integration is such that all poles due to $\Gamma
(a-s)$ and  $\Gamma (b-s)$ lie to the right of the path. Then using (\ref{mellin})
we  have
\be
h_j(x)=\frac{N_j(x)}{2\pi}\;{}_2F_1(\gamma +2j+1,\gamma +2j+1;2j+2;-x)\;.\label{measure}
\ee
This result is valid for $0<s<\gamma +2j+1$ in (\ref{sfh}),
 or $-1<n<\gamma +2j$  in (\ref{coh}), which
contains the compete set of eigenvalues of the number operator ${\bf n}$. 
The integration measure $d\mu_j (\zeta,\bar\zeta ) $ is a product of hypergeometric functions,
and thus the star product (\ref{gsp}) for the fuzzy sphere can now be given as an integral
of hypergeometric functions.

For finite $j$
the coherent states (\ref{trcs}) are not diagonal in ${\bf z}$.
  (${\bf z}$ has only zero eigenvalues for all finite $j$.)  Instead,
\be{\bf  z}| \zeta,j> =\zeta |\zeta,j> -\; \frac{N_j (|\zeta |^2  )^{-\frac12}\;
\zeta^{2j+1}} {\sqrt{(2j)!} \;[f_j(2j)]!     }\;|2j>     \;.\label{zeig}\ee
So as indicated earlier,  we don't have the analogue of (\ref{eefa}).
On the other hand, $|\zeta,j> $  tend to ${\bf z}$ eigenstates in the commutative
limit $j\rightarrow\infty$.
 For this we need
the  asymptotic behavior of $N_j (x  )$.
The result for $x\equiv |\zeta|^2\ll j$ is
[see appendix]
\be
N_j(x)\sim 
(1+x)^{2j}\left(\frac{2j}{1+x}\right)^{2(1-\gamma)}
\exp{\biggl(\frac{1+x}{8j}\biggr) } \label{cloN}\ee
  From this and $\sqrt{(2j)!} \;[f_j(2j)]!     \sim \sqrt{2\pi j}$,
 the last term in (\ref{zeig}) vanishes and so ${\bf z}$ has eigenvalue $\zeta$
in this limit.   We thus expect that $\zeta$ and
$\bar\zeta$  tend to the usual stereographic coordinates $z$ and $\bar z$
of the commutative sphere
in this limit.  This will be demonstrated explicitly later.  Using the asymptotic
expansion of ${}_2F_1$ for large parameters \cite{luke}
\be {}_2F_1 (a_1+2j,a_2+2j;b+2j; -x) \sim
(1+x)^{b-a_1-a_2-2j} (1+O(j^{-1}))\;\;,\quad x\ll j\;,\ee
 and the above
expansion  (\ref{cloN})  for $N_j(x)$ , one can find
behavior of the measure for large $j$ in terms of these variables
\be  d\mu _j(\zeta,\bar\zeta ) \sim \;\frac{j}{\pi}\;\frac{i\;d\zeta \wedge d\bar\zeta
}{ (1+|\zeta |^2  )^2}
 \;\;,\quad x\ll j\;.
\label{measA}
\ee
So we recover the usual measure for $S^2$.  
(We can rescale the coordinates to absorb the $j$ factor.)

From (\ref{cloN}) we can make the observation that if $x$ is `small', more precisely
 of order $1/j$, then 
$N_j(x)$ goes like $2j\; e^y$, where $y$ is a rescaled variable, $x= y/(2j-1)$ .
Small $x$ corresponds to a large radius for the sphere (instead of $1$).  
It thus makes sense that in
this limit we recover the normalization factor associated with standard coherent
states. 

We  now compute some  star products.  For the variables  $\zeta$ and $\bar \zeta$:
\beqa  \bar \zeta \star \zeta &=& |\zeta |^2 \cr
\zeta \star \zeta &=& \zeta ^2 +  \frac{ {\cal Z}_{2j\;2j+2} } {(2j)! [f_j(2j)!]^2}\;, \cr
& & \cr\bar\zeta \star \bar\zeta &=& \bar\zeta ^2 +
 \frac{ {\cal Z}_{2j+2\;2j}  } {(2j)! [f_j(2j)!]^2} \;, \cr
\zeta \star \bar\zeta &=& |\zeta|^2 +  <\zeta|[{\bf z},{\bf z}^\dagger ] |\zeta >
-\frac{ {\cal Z}_{2j+1\;2j+1} } {(2j)! [f_j(2j)!]^2}
\eeqa
The first equation agrees with (\ref{bzsz1}), while the others  contain correction
terms  to (\ref{zsz1}-\ref{zsbs}).  These
correction terms  are due to the fact that the coherent states are not   eigenvectors
of ${\bf z}$, except in the commutative limit.  
Actually, rather than $\zeta$ and $\bar \zeta$, a more usefull set of variables are
 the  `fuzzy' stereographic coordinates $z_F$ and $\bar z_F$
 defined by:
\be  z_F=<\zeta|{\bf z}|\zeta>\;, \qquad  \bar z_F =<\zeta|{\bf z}^\dagger |\zeta> \;.\ee
Using (\ref{zeig}), they are related to  $\zeta$ and
$\bar\zeta$  by
\beqa z_F& =&\zeta - \frac{ {\cal Z}_{2j\;2j+1} }{(2j)! \;([f_j(2j)]!)^2} \;=\;
\sum_{n=0}^{2j-1}  \frac{ {\cal Z}_{n\;n+1} }{n! \;([f_j(n)]!)^2}  \;,   \cr & & \cr
\bar z_F& =&\bar \zeta -
 \frac{ {\cal Z}_{2j+1\;2j} }{(2j)! \;([f_j(2j)]!)^2}\; \;=
\sum_{n=0}^{2j-1}  \frac{ {\cal Z}_{n+1\;n} }{n! \;([f_j(n)]!)^2}   \;,   \eeqa
$ {\cal Z}  _{mn} $ being defined in (\ref{calZ}).  The
transformation from  $\zeta$ and $\bar \zeta$ to  $z_F$ and $\bar z_F$
is nonsingular.  This is because $$ 0< \frac{|\zeta|^{2j}}
{ N( |\zeta|^2)\;(2j)! \;([f_j(2j)]!)^2} \le 1 \;.$$  Also, $z_F$ and $\bar z_F$
tend to $\zeta$ and $\bar \zeta$
in the commutative limit, and just as with the latter variables, the
 star products of $z_F$ and 
$\bar z_F $  reduce
 to the ordinary products in the limit.   For the star commutator of $z_F $ with 
$ \bar z_F $
 we can use the definition
\be
z_F \star \bar z_F - \bar z_F\star z_F
=  <\zeta|[{\bf z},{\bf z}^\dagger ] |\zeta > \ee   For $j\rightarrow \infty$ we can
 replace the
commutator by (\ref{lorhs}) and the right hand side reduces to
$ \frac1{2j} (1+|{z}_F|^2)^2  \;,$
corresponding to the Poisson bracket of $z_F $ and 
$ \bar z_F $ .

For  finite $j$,   ${\cal Z}  _{nm}  $, $n,m=0,1,2,... 2j$ generate
a  $(2j +1)^2$ dimensional algebra given by
(\ref{fnfn}).   This algebra is isomorphic to the algebra of $(2j +1) \times (2j +1)$  matrices
associated with the $j^{\rm th}$ representation of the fuzzy sphere.   The latter are generated
by $(2j +1) \times (2j +1)$  matrix representations for ${\bf z}$ and ${\bf z}^\dagger$ .
 For the case where $j=\frac12$,  the normalization factor is
$N_{\frac12}(x) =(\lambda_{\frac12,-\frac12} + x)/8   \;,  $ $\;\lambda_{j,m}$
given in (\ref{lajm}),
while the  coherent states are
$$|\zeta,\frac12 > = \frac{\lambda_{\frac12,-\frac12}^{\frac12}  |0> \;+\; \zeta |1> }
{\sqrt{\lambda_{\frac12,-\frac12}  + |\zeta|^2}} $$
For the functions ${\cal Z}  _{nm}$ ,  $ n,m=0,1$ in (\ref{calZ}) one gets the
star products
$$
 {\cal Z}  _{n0}  \star {\cal Z}  _{0s}
=\frac8 {\lambda_{\frac12,-\frac12} }  \;{\cal Z}  _{ns} \;,\qquad
 {\cal Z}  _{n1}  \star {\cal Z}  _{1s} = 8   \;{\cal Z}  _{ns} $$
By comparing with (\ref{twod}), we see that this algebra is isomorphic to the matrix algebra
generated by ${\bf z}$ and  ${\bf z}^\dagger $ in the $2\times 2$ representation.
 $z_F$ and $\bar z_F$ can be written as
\be z_F =     \frac {\lambda_{\frac12,-\frac12} }8\; {\cal Z}  _{01} \;,
\qquad \bar z_F =     \frac {\lambda_{\frac12,-\frac12} }8\; {\cal Z}  _{10} \;,\ee
and for their star products we get
\be z_F \star \bar z_F =  \frac {\lambda_{\frac12,-\frac12}   ^2}8{\cal Z}  _{00}\;, \qquad
  \bar z_F \star  z_F =  \frac {\lambda_{\frac12,-\frac12}  }8{\cal Z}  _{11} \;,\ee
along with $ z_F \star  z_F =  \bar z_F \star \bar z_F = 0$.
(This is in contrast with the  ordinary product which gives $|z_F|^2 =
(\lambda_{\frac12,-\frac12}/8)^2\; {\cal Z}  _{00}{\cal Z}  _{11} \;$. )

Finally, we return to the stereographic projection.
For any $j$ we can write it (or more precisely, the inverse  stereographic projection)
in terms of $N_j(x)$.  Here we  invert  (\ref{fzysp})
to solve for the three  dependent
 coordinates $(x_i)_F = <\zeta |{\bf x}_i|\zeta>$ of the fuzzy sphere.
After using
${\bf x}_3 = \alpha ({\bf n} - j)$ ,
 \beqa (x_3)_F &=& \alpha \sum_{n=0}^{2j}
 \frac{n-j }{n! \;([f_j(n)]!)^2}  {\cal Z}_{nn} \cr & &\cr & = &\alpha \biggl[
\zeta \frac{\partial}{\partial\zeta} \ln{ N_j(|\zeta|^2) } - j \biggr]
\;,\label{xf3}\eeqa
while \beqa
(x_1)_F -i(x_2)_F &=& z_F \;\star \;(1 -(x_3)_F \;) \cr & & \cr
&=&  \sum_{n=0}^{2j-1}
 \frac{1-\alpha(n+1-j) }{n! \;([f_j(n)]!)^2}  {\cal Z}_{n\;n+1}\eeqa  $$\qquad
= \;(1+\alpha (j-1))\;\zeta  \;-\;\alpha |\zeta|^2\;
\frac{\partial}{\partial\bar \zeta}\ln{ N_j(|\zeta|^2)}
\; +\; \frac{\alpha(j+1)-1 }{(2j)! \;([f_j(2j)]!)^2} {\cal Z}_{2j\;2j+1}
 $$

\beqa
(x_1)_F +i(x_2)_F &=&  (1 -(x_3)_F \;)\;\star\; \bar z_F \cr & & \cr
&=&  \sum_{n=0}^{2j-1}
 \frac{1-\alpha(n+1-j) }{n! \;([f_j(n)]!)^2}  {\cal Z}_{n+1\;n} \label{xf+}\eeqa
$$\qquad
= \;(1+\alpha (j-1))\;\bar \zeta \; -\;\alpha |\zeta|^2
\frac{\partial}{\partial \zeta}\ln{ N_j(|\zeta|^2)}
 \;+\; \frac{\alpha(j+1)-1 }{(2j)! \;([f_j(2j)]!)^2} {\cal Z}_{2j+1\;2j}
 $$
It identically follows that
\be (x_1)_F\;\star\;  (x_1)_F\;+\;(x_2)_F\;\star\;  (x_2)_F\; +\;
 (x_3)_F\;\star\;  (x_3)_F\;=\;1 \;. \label{fsc}\ee
The right hand sides of (\ref{xf3}-\ref{xf+}) can in principle be reexpressed in terms
of the stereographic coordinates $z_F$ and $\bar z_F$.  For instance, for the case
of $j=\frac12$ we have the  result simple results:  $$
  (x_3)_F =\frac {\bar z_F \star z_F - z_F\star\bar z_F}{2\sqrt{3}(2+\sqrt{3})}  $$
$$ (x_1)_F -i(x_2)_F  = (1-\frac 1{\sqrt{3}})\; z_F \;,\qquad
(x_1)_F +i(x_2)_F  = (1-\frac 1{\sqrt{3}}) \;\bar z_F  $$
We can check that the standard inverse stereographic projection is recovered in the
commutative limit.  Substituting  (\ref{cloN}) into (\ref{xf3}-\ref{xf+}) gives
$$ (x_3)_F \rightarrow \frac{|\zeta|^2 - 1}{|\zeta|^2 + 1}   $$
$$ (x_1)_F -i(x_2)_F  \rightarrow \frac{2 \zeta}{|\zeta|^2 + 1}  \;,   \qquad
(x_1)_F +i(x_2)_F   \rightarrow \frac{2 \bar\zeta}{|\zeta|^2 + 1}     \;.  $$ Since
 ($\zeta ,\bar\zeta$) and $(z_F,\bar z_F)$ coincide in the limit they both reduce to
 the usual stereographic coordinates $(z,\bar z)$
of the commutative sphere
in this limit.  (A different deformation of the standard stereographic projection can be
found in \cite{HA}.)

\bigskip
\noindent
{\bf Acknowledgement}
We are gratefull for valuable discussions with A. P. Balachandran, G. Karatheodoris,
 D. O'Connor and  C. Zachos (who also informed us of numerous references).
This work was supported by the joint NSF-CONACyT grant E120.0462/2000.  G. A. was
supported in part by the U.S. Department of Energy
 under contract number DE-FG02-85ER40231, and 
A. P. and A. S. were supported in part by the U.S. Department of Energy
 under contract number DE-FG05-84ER40141.

\bigskip
\noindent
\appendice

Here we obtain the expression (\ref{ceffj}) for $N_j(x)$ for the fuzzy sphere,
and give its asymptotic expansion (\ref{cloN}).

From (\ref{fjn}),
\be \;([f_j(n)]!)^2  = \frac{(2j+1)! \;\;\Gamma(\sqrt{j(j+1)} + j  -  n)\;{}^2}
{(2j-n)!\; \;\Gamma(\sqrt{j(j+1)} + j +1)\;{}^2} \;.   \ee
Substituting into (\ref{Nsj}) gives
\be N_j (x  )  = \frac {\Gamma(\gamma + 2j +1)\;{}^2}{\Gamma(2j +2)} \;\;
 \sum_{n=0}^{2j}\;\frac {\Gamma(2j-n +1)}{\Gamma(\sqrt{j(j+1)} + j - n)\;{}^2 }
\; \frac{x^n} {n!} \;\;,\ee where $\gamma = \sqrt{j(j+1)} - j $.
 Now replace the summation index by $m=2j-n$ to get
\be N_j (x  )  = \frac {\Gamma(\gamma + 2 j +1)\;{}^2}{\Gamma(2j +2)} \;x^{2j}\;
 \sum_{m=0}^{2j}\;\frac {\Gamma(m+1)\;{}^2}{\Gamma(2j+1-m)\;\Gamma(\gamma + m)\;{}^2 }
\; \frac{x^{-m}} {m!} \;\;.\ee  From the
identity $$\frac{\Gamma(a+1)}{\Gamma(a-m)} = - (-1)^m
\frac{\Gamma(m+1-a)}{\Gamma(-a)}  \;,$$ it follows that
\be   N_j (x  )  =\;-\; \frac {\Gamma(\gamma + 2j +1)\;{}^2}{\Gamma(-2j-1)\;
\Gamma(2j +2)\;{}^2} \;x^{2j}\;
 \sum_{m=0}^{2j}\;\frac {\Gamma(m-2j)\;\Gamma(m+1)\;{}^2}{\Gamma(\gamma + m)\;{}^2 }
\; \frac{(-x)^{-m}} {m!} \;\;.\label{arfnj}
\ee  This is just (\ref{ceffj}), once we use the definition
$$ {}_3F_2 (a_1,a_2,a_3;b_1,b_2; x) = \frac{\Gamma(b_1)\Gamma(b_2)}
{\Gamma(a_1)\Gamma(a_2)\Gamma(a_3)} \;\sum_{n=0}^{\infty}
\;\frac {\Gamma(a_1+n)\Gamma(a_2+n)\Gamma(a_3+n)} {\Gamma(b_1+n)\Gamma(b_2+n)}\;
 \frac{x^n} {n!} \;\;.$$ \be\label{hf}\ee  
The infinite series expression is valid provided no $a_i$ are negative
integers, which excludes our case.  On the other hand,
 the series terminates when $a_i$ is a negative
integer, say $-2j$ (our case), at $n=2j$ , and the hypergeometric function (\ref{hf})
is then referred to as  an extended
 Laguerre polynomial.  To find the asymptotic representation for large $j$
 one can use the Darboux analysis \cite{luke}. For the extended
 Laguerre polynomials the result up to $1/j^2$ corrections is
\beqa
{}_3F_2 (a_1,a_2,-2j;b_1,b_2; -x) \sim
\frac{\Gamma(b_1)\Gamma(b_2)}{\Gamma(a_1)\Gamma(a_2)}(1+x)^{2j}\left(\frac{2jx}{1+x}
\right)^\beta
\times\\ \nonumber
\exp\left\{-\frac{u}{2jx}+\frac{v}{2j}+O(j^{-2})\right\}+O(j^{-2})\;\;,\label{asympt}
\eeqa
where $0<x\ll 2j$ and
\beqa
\beta & =& a_1+a_2-b_1-b_2\;,\cr
 u &= &a_1a_2 - b_1b_2 +\beta
(1-a_1-a_2)\;\;,\cr
v& =& - a_1a_2 + b_1b_2 +\frac{1}{2}\beta (a_1+a_2+b_1+b_2-1)\;\;.
\eeqa
Using $\beta = 2(1-\gamma)\;,\;u=-(1-\gamma)^2\;$ and $v=\gamma (1-\gamma)$, we get
(\ref{cloN}).

\end{document}